\newcommand{\be}{\begin{eqnarray}}
\newcommand{\ee}{\end{eqnarray}}
\newcommand{\dd}{{\rm d}}

\documentstyle[12pt]{article}

\begin{document}

\begin{flushright}
BNL-HET-98/2 \\
TTP 97-19\footnote{Talk given at the {\em Workshop on Physics at the
First Muon Collider and at the Front End of a Muon Collider},
Fermilab, November 6--9, 1997.}\\ hep-ph/9801218\\ January 1998
\end{flushright}

\vspace*{3mm}

\begin{center}
{\LARGE \bf Coherent muon--electron conversion\\[2mm] in muonic atoms}
\vspace*{5mm}

{\large \bf Andrzej Czarnecki and William J.\ Marciano}

\vspace*{2mm}

{\em Brookhaven National Laboratory, Upton, New York 11973}

\vspace*{4mm}

{\large \bf Kirill Melnikov}

\vspace*{2mm}

{\em Institut f\"ur Theoretische Teilchenphysik, \\
Universit\"at Karlsruhe,
D-76128 Karlsruhe, Germany}
\end{center}

\begin{abstract}
Transition rates for  coherent muon-electron conversion in muonic
atoms, $\mu N \to eN$, are computed for various types of muon number
violating interactions.  Attention is paid to relativistic atomic
effects, Coulomb distortion, finite nuclear size, and nucleon
distributions.  Discrepancies with previously published results are
pointed out and explained.  Results are presented for several elements
of current and future experimental interest.
\end{abstract}

\section {Introduction}

In the standard model (SM) of electroweak interactions, leptons of
different flavors do not mix because of vanishing neutrino masses.  On
the other hand, lepton flavor violation is predicted in
various extensions of the SM.  Hence, its observation would be direct
evidence of physics beyond the SM.  

Because the muon is a relatively
stable particle which can be abundantly produced, muon--number
violating processes are of particular  interest.  Most
recently, the following processes have been studied experimentally
(for a review see e.g.~\cite{Schaff93,balholm}): $\mu \to e \gamma$,
muonium-antimuonium oscillations, and, the subject of the present
work, muon--electron conversion in the field of a nucleus.  Muon
number violation has also been searched for in kaon, tau, and $Z$
decays.

A particularly sensitive search for muon number violation is the
coherent conversion in a muonic atom, formed by muon and a
nucleus. Here by coherent conversion we understand a process $\mu ^- N
\to e^- N $ in which the nucleus $N$ remains in its initial state (up
to recoil effects).  The rate of the coherent conversion is enhanced
with respect to processes with a nuclear excitation by a factor of the
order of the number of nucleons.

Since the original pioneering papers on the conversion theory
\cite{wf59,Cabibbo59,rosen60}, there have been several theoretical
efforts intended to provide an 
accurate description of  coherent muon conversion.
There are two kinds of theoretical issues: first, the short
distance effects which are responsible for the muon number violation
(presumably caused  by some ``new physics'');
second, the long distance atomic physics of the muonic atom, which
also determines the transition rate.

The first group of problems has been studied in many extensions of the
standard model (for a review see
\cite{MarcianoOrbis,Kosmas:1994pt,Vergados:1986pq}).  In particular,
in ref.~\cite{Marciano77PRL} the rate of the coherent conversion $\mu
^-N \to e^-N $ was calculated in a variety of gauge models.  It was
pointed out in that paper that in a large class of models the
conversion can be much more probable than the decay $\mu\to e\gamma$.
For example, there can be logarithmic enhancements of the form factors
leading to conversion which are absent in the decay rate.  Such
logarithmic effects were also recently discussed in
\cite{Santamaria97}.

Weinberg and Feinberg \cite{wf59} focused on an electromagnetic
mechanism of transferring the energy yield to the nucleus.  The
structure of this electromagnetic interaction is richer than for the
$\mu\to e\gamma$ decay since the photon need not be on mass shell.
Therefore, it is possible that conversion can occur even if $\mu\to
e\gamma$ is forbidden for some reason.  The matrix element for
conversion contains monopole terms which do not contribute to the
decay $\mu\to e\gamma$, because the longitudinal polarization states
are possible only for virtual photons.  In addition, there may be
conversion amplitudes other than photon mediated processes.
This makes the conversion on nuclei a particularly interesting process
to study.

The early theoretical studies of  muon conversion into electrons on
nuclei performed in \cite{wf59,Cabibbo59,rosen60} are valid mainly for
conversion on light nuclei.  The developments before the year 1978
have been summarized in \cite{MarcianoOrbis}.  In heavier atoms new
effects become important: relativistic components of the muon wave
function, Coulomb distortion of the outgoing electron, and the finite
nuclear size.  These were first addressed in \cite{Shanker79}.  Nuclear
effects were also analyzed, albeit in a non-relativistic
approximation, in \cite{Chiang93} and, more recently, in
\cite{Kosmas:1997a,Kosmas:1997b}.

Recently, we have undertaken a new calculation of the full
relativistic atomic physics aspect of the conversion.  In this talk
our main results are summarized.  The details of the calculation and a
more extensive analysis will be given in a forthcoming paper
\cite{CMMinprep}.

\section{General description}

The muon--electron transition can occur via various mechanisms (see
e.g.~\cite{MarcianoOrbis,BNT,Barbieri:1995}).
To keep the description fairly general, it is convenient
to write down a low energy  
effective Hamiltonian for the $\mu \to e$ transition:
\be
{\cal H} &=& - \bar e \hat O \mu + h.c.,
\\
\hat O &=& -\sqrt {4\pi \alpha }
\left[ \gamma_\alpha \left(f_{E0} - f_{M0}\gamma_5\right)
\frac {q^2}{m^2}
+i\sigma_{\alpha \beta} {q^\beta \over m} 
\left(f_{M1} + f_{E1}\gamma_5\right)
 \right]A^\alpha (q)
\\ \nonumber
&+&{G_F\over \sqrt{2}} \gamma_\alpha (a-b\gamma_5)J^\alpha,
\\
J^\alpha &=& \bar u\gamma^\alpha u  + c_d \bar d\gamma^\alpha d.
\ee
In this equation  $m=m_\mu$ is muon mass and
$a,b,f_{E0,1},f_{M0,1}$ are dimensionless coupling constants.

The part of the Hamiltonian containing the photon
field $A_{\alpha}$ describes the transition of a muon into an
electron and an on--shell photon, $\mu \to e \gamma$, whose  rate
is
$\Gamma(\mu\to e\gamma) = \alpha m\left( 
|f_{M1}|^2 + |f_{E1}|^2 \right)/2$,
which gives the   branching ratio
$Br(\mu\to e\gamma) = { 
\Gamma(\mu\to e\gamma)
/
\Gamma(\mu\to e\bar \nu_e \nu_\mu)}
={ 96\pi^3\alpha }
\left( |f_{M1}|^2 + |f_{E1}|^2 \right)/( G_F^2m^4)$.
Due to gauge invariance, the transition vector and axial currents
(terms proportional to $f_{E0}, f_{M0}$) cannot contribute to the
transition to an on--shell photon \cite {wf59}. We explicitly account
for this by factoring out $q^2/m^2$.  We further assume that all
effective coupling constants introduced in the Hamiltonian are slowly
varying functions of all external parameters (photon off--shellness,
muon off--shellness etc.) and therefore can be considered as constants
in the course of the calculation.

To calculate the transition rate for the coherent conversion $\mu N
\to e N$ we first average the effective Hamiltonian
over the nucleus.  The result of this average 
depends on muon and electron wave functions and the field of the
nucleus:
\be
H_{\rm int} = \langle N | {\cal H} | N \rangle ,\qquad\langle N | N
\rangle =1. 
\label {norm}
\ee

To calculate a matrix element of an arbitrary operator
$\hat Q$ between two nuclei, we use the following approximation:
$$
\langle N | {\hat Q} | N \rangle = \int {\rm d}^3 r 
\Bigg (Z \rho _p(r) \langle p |Q(r)|p \rangle +
(A-Z) \rho _n(r) \langle n |Q(r)|n \rangle \Bigg ).
$$

In the above equation $Z$ is the number of protons in the nucleus $N$
and $(A-Z)$ is the number of neutrons. Also, $|p\rangle$ and $| n
\rangle$ denote states of a single proton and neutron respectively,
with densities normalized as follows: 
\be
\int {\rm d}^3r \rho _p(r) =1,\qquad\int {\rm d}^3r \rho _n(r) =1.
\ee

To obtain matrix elements of the Hamiltonian, one should calculate the
matrix element of the quark currents between two nucleons. A typical
momentum transfered by the current is of the order of the muon mass, a
small quantity compared to the mass of the nucleus.  The time
component of the soft current counts the number of constituent quarks
in the nucleon. We therefore obtain
\be
\langle p | \bar u\gamma^0 u  + c_d \bar d\gamma^0 d 
|p \rangle &=& 2 + c_d,
\nonumber \\
\langle n | \bar u\gamma^0 u  + c_d \bar d\gamma^0 d 
|n \rangle &=& 1 + 2c_d.
\ee
The matrix element of  spatial components of the current is
proportional to the velocities of the constituents and is
negligible in the present  problem.

Calculating the matrix element of the Hamiltonian with respect to the
nucleus states, we arrive at an effective Hamiltonian for 
the coherent $\mu \to e$ conversion in the field of the nucleus:
\be
H_{\rm int} &=& H_1+H_2+H_3,
\nonumber \\
H_1 &=&
e \int {\rm d}^3r \overline{\Psi} _e (r) \gamma _0 
 \left(f_{E0} - f_{M0}\gamma_5\right) \Psi _\mu (r) 4\pi Z \rho _p (r)
\nonumber \\
H_2 &=& 
e \int {\rm d}^3r \overline \Psi _e (r)i\sigma_{\alpha \beta}  
\left(f_{M1} + f_{E1}\gamma_5\right) \Psi _\mu (r) F_{\alpha \beta}(r)
\nonumber \\
H_3 &=&{G_F\over \sqrt{2}} 
 \int {\rm d}^3r 
\overline \Psi _e (r)  \gamma_0 (a-b\gamma_5) \Psi _\mu (r)
\nonumber \\ 
&& \times
 \Bigg (Z (2+c_d)\rho _p(r) +(A-Z)(1+2c_d) \rho _n (r) \Bigg )
\ee

In the above equations the fields $\Psi _e (r) $ and $\Psi _\mu (r)$
stand for the second quantized operators, and the electromagnetic
tensor $F _{\alpha \beta}$ is understood as a classical electric field
produced by the nucleus.  The matrix element of $H_{\rm int}$ taken
between appropriate initial and final states will give the amplitude
for the coherent conversion.

In our case the initial state is the muon in the $1S$ orbit around the
nucleus; the final state is an electron with the energy equal to the
energy of the initial muon.  The corrections due to the nucleus recoil
are small and we do not consider them here.

In their pioneering work on muon--electron conversion \cite {wf59}
Weinberg and Feinberg performed an approximate calculation of the
coherent conversion rate.  Using simple estimates, we would like to
show when complete treatment of the problem requires going beyond the
approximations used in Ref.~\cite{wf59}.  For this purpose we describe
the scales relevant for the problem.  The wave function of the muon
bound in the lowest orbit is characterized by the Bohr radius $a _B =
(\alpha Z m)^{-1}$.  The radius of the nucleus $R_N$ scales like
$m_{\mu} R_N \sim (Z/4)^{1/3}$.  Evidently, the nucleus can be
considered as point--like only if the Bohr radius is much larger than
the radius of the nucleus $a_B \gg R_N$.  This implies $Z
\ll 60$.  On the other hand, the relativistic corrections are governed
by the parameter $Z \alpha$. Therefore, for high $Z$ elements such as ${\rm
Pb}$, it is not clear a priori if the non-relativistic treatment of
the muon bound state is sufficient.  Moreover, another physical effect
is governed by the same parameter $Z\alpha$.  Consider the case when
$\mu \to e $ conversion occurs due to an exchange of the photon with
the nucleus. In this case the kinematics of the decay dictates that
the virtuality of the photon is determined by the mass of the decaying
muon. The process can be considered point--like, if this
scale is much less than the Bohr radius. This implies that the
photon--mediated conversion can be considered as a point--like process
for $Z\alpha \ll 1$. Therefore, for heavy nuclei such as ${\rm Pb}$ the
consideration of the conversion process as point like is no longer
valid.  The appearance of the part $H_2$ in the effective Hamiltonian
$H_{\rm int}$ reflects this observation: one notes that the $\mu \to
e$ transition current couples to the electric field of the nucleus, 
not to the proton density directly.

Clearly, for light nuclei one can rely on the hierarchy of the scales
and perform an approximate calculation of the rate.  In this case the
conversion rate will be proportional to the square of the muon wave
function at the origin and the square of the nucleus form-factor.
Such an approximation was used in Ref. \cite {wf59}.  When the charge
of the nucleus grows, all scales relevant to the problem become
comparable and the above approximations cannot be trusted. In
this case a correct treatment of the problem requires solving the Dirac
equation for both muon and electron wave functions in the field of the
nucleus.

To describe proton and neutron distributions in the nucleus we use
two--parametric Fermi functions:
\be
\rho _{p(n)} (r) = \frac {\rho _0}{1+e^{(r-r_{p(n)})/a _{p(n)}}},
\qquad  \int d^3r \rho _{p(n)} (r) =1.
\label {Fermi}
\ee
In the above equation $r_{p(n)} \sim A^{1/3}$ is the radius of the
proton (neutron) fraction of the nucleus and $a_{p(n)}$ is the thickness
of the boundary of the proton (neutron) fraction.  Precise values of
these parameters depend on the nucleus and can be found in tables
\cite{vries87,garcia92}.

The transition rate for the coherent conversion is given by:
\be
\omega (\mu ^- N \to e^- N) = \omega _{conv} = \sum _{\lambda _f}
|H_{\rm int}^{fi}|^2.
\ee
Here the sum goes over all quantum numbers which the electron in the
final state can have in addition to energy. 
The muon bound state wave function
is normalized to unity. The electron wave function of the continuous
spectrum
is normalized such that:
\be
\int {\rm d}^3r \psi ^* _{E',\lambda '}(r)\psi _{E,\lambda} (r) = 
\delta _{\lambda,\lambda '}2\pi \delta (E'-E)
\ee
where $\lambda $ denotes a set of all discrete quantum numbers which
electron obeys in addition to energy.

It is convenient to use the expression for the wave functions as suggested
in \cite {RelElectron}:
\be
\psi ^{m}_{k} =
\left ( \begin{array}{c}
g^{(k)}(r)\chi ^{m}_{k}\\
if^{(k)}(r)\chi ^{m} _{-k}
\end{array} \right ).
\ee
In this equation $m$ is the eigenvalue of the operator $J_z$, $J =
L+S$; $-k$ is the eigenvalue of the operator $ {\boldmath L}\cdot {\boldmath
\sigma} +1$ (see e.g.\ \cite{RelElectron}); $\chi ^{m}_{k}$ is an
orthonormalized spinor.  For the muon wave function we use the lowest
energy bound state wave function; this suggests that $j=1/2$, $m = \pm
1/2$, $k=-1$. We will distinguish muon and electron radial wave
functions by a subscript.

A calculation of the explicit expression
for the transition rate $\omega _{conv}$ is now straightforward. 
We find
\be
\omega _{conv}&=&
\left| \int \dd r r^2 \left[
 \tilde f_{E0}(r) ( g^-_e g^-_\mu + f^-_e f^-_\mu )
 +{f_{M1}\over m} (g^-_e f^-_\mu + f^-_e g^-_\mu) {dV\over dr}
                 \right] \right|^2
\nonumber \\
&+&
\left| \int \dd r r^2 \left[
 \tilde f_{M0}(r) ( g^-_e g^-_\mu + f^-_e f^-_\mu )
 +{f_{E1}\over m} (g^-_e f^-_\mu + f^-_e g^-_\mu) {dV\over dr}
                 \right] \right|^2,
\nonumber \\
\tilde f_{E0}(r)&=& -f_{E0} \frac {4\pi Z\alpha}{m^2} \rho _p (r)
 + {G_F\over \sqrt{2}}a \left[Z(2+c_d)\rho _p (r) + N(1+2c_d)\rho _n
(r)\right], 
\nonumber \\
\tilde f_{M0}(r)&=& -f_{M0} \frac {4\pi Z\alpha}{m^2} \rho _p (r)
 + {G_F\over \sqrt{2}}b \left[Z(2+c_d)\rho_p (r) + N(1+2c_d)\rho_n
(r)\right], 
\label{eq:rateGen}
\end{eqnarray}
where $V(r) = -eA^0(r)$ is the muon potential energy in the field
of the nucleus.

The above equation provides a general expression for the rate of the
reaction $\mu ^- N \to e^- N$ which we will use for numerical analysis
in the next section. The radial wave functions $f_{e,\mu}^-$ and
$g_{e,\mu}^-$ are obtained by solving the Dirac equation.  Inside the
nucleus and close to it these solution are found numerically.  At large
distances we match the numerical solutions to the exact Coulomb
wave functions.

\section {Numerical analysis}
\label{sec:integrals}
In general, the following integrals are needed for the description of
the transition rate:
\be
I_1 ^p&=& -{4\pi Z\alpha\over m^2} \int \dd r r^2 \rho _p(r) g_\mu^-
g_e^-,  \qquad
I_2^p =-{4\pi Z\alpha\over m^2} \int \dd r r^2 \rho _p(r) f_\mu^- f_e^-
\nonumber \\
I_1^n&=& -{4\pi Z\alpha\over m^2} \int \dd r r^2 \rho _n(r) g_\mu^-
g_e^-, \qquad
I_2^n =-{4\pi Z\alpha\over m^2} \int \dd r r^2 \rho _n(r) f_\mu^- f_e^-
\nonumber \\
I_3&=& {1\over m} \int \dd r r^2 {\dd V(r)\over \dd r} g_\mu^- f_e^-,
\qquad
I_4= {1\over m} \int \dd r r^2 {\dd V(r)\over \dd r} f_\mu^- g_e^-.
\ee
Tables with the values of these integrals for various elements will be
given in \cite{CMMinprep}.  Using those values, the  conversion rate is
\be
\omega _{conv} &=& 3 \cdot 10^{23} 
(\omega _{conv}^{(1)} + \omega _{conv}^{(2)})\; {\rm sec}^{-1} 
\label {numerics}
\\[1.5mm]
\omega _{conv}^{(1)} &=& \left | f_{E0} I_p -
\frac {G_F}{\sqrt {2}} \frac {m^2}{4\pi Z \alpha}a \Big (Z(2+c_d)I_p
+N(1+2c_d) I_n \Big )+ f_{M1} I_{34} \right | ^2,
\nonumber \\
\omega _{conv}^{(2)} &=& \left | f_{M0} I_p -
\frac {G_F}{\sqrt {2}} \frac {m^2}{4\pi Z \alpha} b \Big (Z(2+c_d)I_p
+N(1+2c_d) I_n \Big )+ f_{E1} I_{34} \right | ^2,
\ee
where
\be
I_p = -\Big ( I_1^p+I_2^p \Big ),
\qquad I_n = -\Big ( I_1^n+I_2^n \Big ),
\qquad I_{34} = I_3 +I_4.
\label {ip}
\ee
All dimensional parameters in the above equations
are expressed in fermi. 
For  proton and neutron distributions we use the two--parameter 
Fermi distribution (\ref {Fermi})
with parameters taken from  Ref.  \cite {vries87,garcia92}.

For the present application we neglect the effects of the 
vacuum polarization; such approximation is justified in
view of much bigger errors in wave function integrals induced directly
by the nuclear distribution uncertainties.

\section {Branching ratio for $\mu$-$e$ conversion in 
a specific model}
\label{sec:mechan}

We consider here one specific model which predicts lepton flavor
violation, a supersymmetric grand-unified theory discussed in
Ref. \cite {Barbieri:1995}.
 
The analysis of Ref. \cite {Barbieri:1995}
implies that magnetic couplings $f_{M1}$ and $f_{E1}$ (cf. Eq.(2)) are
significantly enhanced in a large region of the parameter space of
those models in comparison with other  couplings in the
effective Hamiltonian.  Therefore, it is reasonable to neglect all
other couplings in the effective Hamiltonian and to analyze the
dependence of $\omega _{conv} /\omega _{capt}$ on the choice of
target.

With these approximations the ratio of the conversion rate
to the capture rate becomes
\be
\frac {\omega _{conv}}{\omega _{capt}} &=&  3 \cdot 10^{12} 
\Big ( |f_{E1}|^2+|f_{M1}|^2 \Big ) B(A,Z),
\label{eq:rate}
\\
\mbox{with} &&
B(A,Z) = 10^{11}~\frac {I_{34}^2}  {\omega_{capt}/{\rm sec}^{-1}}.
\ee

We compare the above formula with the Weinberg--Feinberg
approximation, in which the rate is obtained by replacing $B(A,Z)$ in
(\ref{eq:rate}) by $B_{WF}(A,Z)$:
\be
B_{WF}(A,Z) = 8\alpha^5 m Z_{\mbox{eff}}^4 Z F_p^2 
    {1\over \omega_{capt}}  {1\over 3\cdot 10^{12}}
\ee
In this formula $Z_{\mbox{eff}}$ denotes the ``effective $Z$,'' obtained by
averaging the muon wave function over the nuclear density (see e.g.\
\cite{Chiang93}), and $F_p$
is the formfactor describing the charge distribution, given by
\begin{equation}
F_p = {\int \dd^3 r \rho(r) {\sin m r\over m r} \over
\int \dd^3 r \rho(r)}.
\label{eq:ff}
\end{equation}
In table \ref{tab:capture} we show the $B$ and $B_{WF}$ coefficients
for three elements.  We conclude that in this particular model the
ratio of the coherent conversion rate to the capture rate, as
described by our $B(A,Z)$, does not change significantly with changing
the target.  The WF approximation tends to slightly overestimate the
conversion rate.  
In ref.~\cite{Shanker79} various corrections to the WF approximations
have been studied; in that approach $B_{WF}$ should be replaced by
$B_S \equiv C_1C_2C_3 B_{WF}$, where the correction factors
$C_{1,2,3}$ are listed in  table III in \cite{Shanker79}.  We agree
with these results for non-photonic mechanisms of conversion.
However, for the
photonic case we are considering here, our results differ from
\cite{Shanker79} even stronger than from the WF approximation, as can
be seen in table \ref{tab:capture}.  This is because the effect of
the non-locality in the interaction with the electric field was not
taken into account in \cite{Shanker79}.  It is especially important
for heavy elements.
\begin{table}[ht]
  \caption{Comparison of our results for the coefficient $B(A,Z)$ in
    the rate formula (\protect\ref{eq:rate}) with the 
    Weinberg-Feinberg approximation $B_{WF}(A,Z)$ and with $B_S(A,Z)$
    obtained from ref.~\protect\cite{Shanker79}.  $Z_{\mbox{eff}}$
    is taken from \protect\cite{Chiang93} and the capture rates
    $\omega_{capt}$ from \protect\cite{capture}; $F_p$ is computed
    using (\protect\ref{eq:ff}).}
\vspace*{2mm}
  \label{tab:capture}
\begin{center}
    \leavevmode
    \begin{tabular}{clccccr}
\hline
 Element &  $B(A,Z)$& $B_{WF}(A,Z)$ & $B_S(A,Z)$ & $Z_{\mbox{eff}}$ & $F_p$ & 
              $\omega_{capt},~[10^6/{\rm sec}]$ \\[1mm]
\hline
Al & 1.1(1) & 1.2 & 1.3 & 11.62 & 0.63&  0.7   \\
Ti & 1.8 & 2.0 & 2.2 & 17.61 &0.53 & 2.6\\
Pb & 1.25(15) & 1.6 & 2.2 & 33.81 & 0.15 & 13.0\\
\hline
    \end{tabular}
\end{center}
\end{table}

It is  instructive to compare the conversion rate to the branching ratio
for $\mu \to e\gamma$ in this model:
\be
{Br(\mu\to e\gamma)\over \omega_{conv}/\omega_{capt} }
=
{96\pi^3\alpha\over G_F^2 m^4} {1\over 3\cdot 10^{12} B(A,Z)}
\approx {428 \over B(A,Z)}. 
\ee
This ratio varies from 389 for $^{27}{\rm Al}$ to 
238 for $^{48}{\rm Ti}$, and increases again for heavy elements to 342
for $^{208}{\rm Pb}$.

\section{Bounds on the muon number violating couplings from SINDRUM}
\label{sec:bounds}

Using updated results for the atomic part of the theoretical
description of the coherent muon electron conversion, we reconsider
the analysis performed in Ref. \cite {Honecker:1996}.  In that paper
the upper limit for the ratio $\omega _{conv} / \omega _{capt}$ has
been reported based on the measurement using $^{208}{\rm Pb}$.  These
results were combined with previous measurement on $^{48}{\rm Ti}$
target to obtain the bounds on muon--number violating coupling
constants.  In that analysis it was assumed that the effective
Hamiltonian does not contain a piece which corresponds to the photon
mediated conversion.

As a first step we have to switch to the model and notations of
Ref.\ \cite {Honecker:1996}. The model, considered in Ref. \cite
{Honecker:1996}, corresponds to the exchange of a heavy vector or
scalar boson which mediates $\mu \to e$ transition.  For this reason,
we should equate $f_{E0,1}$ and $f_{M0,1}$ to zero in our general
expression for conversion rate, eq.~(\ref{eq:rateGen}). 
Also, to switch to their notations,
one should substitute $b=a$,
$ac_d \to  ({g_V^{0} - g_V^{1}})/{2}$, and
$a \to (g_V^{0}+g_V^{1})/{2}$
in that equation.
Constants  $g_V^{0}, g_V^{1}$ then parameterize the coupling of
the heavy gauge boson to isoscalar and isovector parts of the quark
current respectively.

Then, one gets for the conversion rate:
\be
\omega _{conv} \sim I_p^2 \Bigg [ g^0 _V \Bigg (1-\frac {N}{A}\xi \Bigg)
+g^1_V \Bigg (\frac {(Z-N)}{3A}+\frac {N}{3A}\xi \Bigg ) \Bigg ]^2.
\ee
In this equation we use the notation (cf. Eq. (\ref {ip})):
$\xi = { I_p}/{ I_n}$.
Using the bounds on the $\mu \to e$ branching ratios measured with
Ti and Pb targets, we derive the new bounds for the coupling
constants $g_V^{0},~~g_V^{1}$:
\be
|g^0_V| < 8\pm 2 \cdot 10^{-7},\qquad 
|g^1_V| < 40 \pm 13 \cdot 10^{-6}.
\label {newbounds}
\ee

The variation in the boundaries shown above is the estimate 
of the uncertainty of the result induced by uncertainties in the
input nuclear parameters.

These new bounds should be compared with the values
\be
|g^0_V| < 3.9 \cdot 10^{-7},
\qquad 
|g^1_V| < 9.7 \cdot 10^{-6}.
\label {oldbounds}
\ee
which were given in Ref.\ \cite {Honecker:1996} 
using the results of Ref.\ \cite {Shanker79}.

There are two reasons for such large differences.  As we have
mentioned already, in Ref. \cite {Honecker:1996} the results of
Ref. \cite {Shanker79} have been used to interpret experimental data.
We note in this respect, that Table I of ref. \cite {Shanker79} does
not provide correct results -- the results for $\omega _{conv}/ \omega
_{capture}$ quoted there are about a factor of 2 too large.  Second,
in the case of a Pb target the difference in proton and nucleon
distributions becomes quite noticeable. However, the numbers in Table
I of \cite {Shanker79} were obtained using identical proton and
neutron distributions.  These two reasons conspire to give quite a
large discrepancy.

We note, however, that the values for the bounds quoted above
(especially for $g^1_V$) are very sensitive to the input parameters.
This happens, because the bounds are derived from a system of linear
equations in which the coefficients of $g^1_V$ there are much smaller
than all other parameters. Therefore even a small variation in the
input parameters can generate quite substantial change in the
resulting value for $g^1_V$.

\section{Acknowledgement}
We are grateful to Dr.~P.~Goudsmit for providing us with
ref.~\cite{Engfer74} and to Dr.~Roland Rosenfelder for discussions and
providing his Fortran codes for comparisons.  A.C.~thanks
Dr.~T.~Koz{\l}owski for explaining some details of the Sindrum II
experiment.  This research was supported by BMBF-057KA92P; by
``Graduiertenkolleg Elementarteilchenphysik'' at the University of
Karlsruhe; and by U.S.~Department of Energy under contract
number DE-AC02-76CH00016.

\end{document}